\documentclass[prl,nofootinbib,twocolumn,superscriptaddress,preprintnumbers,balancelastpage,longbibliography]{revtex4-1}

\usepackage{graphicx}
\usepackage{color}
\usepackage{hepunits}

\usepackage[english]{babel}

\newcommand{\es}[2] {\begin{equation} \label{#1} \begin{split} #2 \end{split} \end{equation}}

\newcommand{\be}{\begin{equation}}
\newcommand{\ee}{\end{equation}}

\newcommand{\m}{{\bf \hat m}}

\newcommand{\g}{g_{a \gamma \gamma}}

\begin{document}
\title{Radio Signals from Axion Dark Matter Conversion \\
in Neutron Star Magnetospheres }

\author{Anson Hook}
\email{hook@umd.edu}
\affiliation{Maryland Center for Fundamental Physics, Department of Physics, University of Maryland, College Park, MD 20742.}

\author{Yonatan Kahn}
\email{ykahn@princeton.edu}
\affiliation{Department of Physics, Princeton University, Princeton, NJ 08544, USA}

\author{Benjamin R. Safdi}
\email{bsafdi@umich.edu}
\affiliation{Leinweber Center for Theoretical Physics, Department of Physics, University of Michigan, Ann Arbor, MI 48109 U.S.A.}

\author{Zhiquan Sun}
\email{zqsun@umich.edu}
\affiliation{Leinweber Center for Theoretical Physics, Department of Physics, University of Michigan, Ann Arbor, MI 48109 U.S.A.}

\preprint{LCTP-18-09, PUPT-2558}

\date{\today}

\begin{abstract}
We show that axion dark matter (DM) may be detectable through narrow radio lines emitted from neutron stars.
The neutron star magnetosphere hosts a strong magnetic field and a plasma frequency that increases towards the neutron star surface.  As the axions pass through the magnetosphere, they can resonantly convert into radio photons in a narrow region around the radius at which the plasma frequency equals the axion mass. The bandwidth of the signal is set by the small DM velocity dispersion far away from the neutron star. 
We solve the axion-photon mixing equations, including a full treatment of the magnetized plasma and associated anisotropic dielectric tensor, to obtain the conversion probability. We discuss possible neutron-star targets 
and how they may probe the QCD axion parameter space in the mass range of $\sim$$0.2$-40 $\mu$eV.
\end{abstract}

\maketitle

The QCD axion is one of the best-motivated dark matter (DM) candidates: in addition to explaining the observed abundance of DM~\cite{Preskill:1982cy,Abbott:1982af,Dine:1982ah}, the axion may also resolve the strong {\it CP} problem~\cite{Peccei:1977hh,Peccei:1977ur,Weinberg:1977ma,Wilczek:1977pj}.  However, testing the axion DM hypothesis is notoriously difficult, due to the fact that the axion is predicted to interact only very weakly with ordinary matter. In particular, axions interact with photons through ${\mathcal L}_{a \gamma} = -\frac{1}{4}g_{a \gamma \gamma} a {\bf E} \cdot {\bf B}$, where $g_{a\gamma\gamma} \sim 10^{-15} \ \GeV^{-1}$ for an axion of mass $m_a = 10^{-6} \ \eV$. Direct detection experiments such as ADMX~\cite{Asztalos:2001tf,PhysRevLett.104.041301,Shokair:2014rna} are just beginning to probe small regions of the QCD axion parameter space by exploiting this small coupling between the axion and electromagnetism.  In this Letter, we 
discuss a new avenue for testing the axion DM hypothesis: axion DM indirect detection.  We show that in the presence of axion DM, monochromatic radio signals are emitted from neutron stars (NSs) due to axion-photon conversion within the high magnetic field regions in the NS magnetosphere.  

As we point out in this Letter, and was noted in~\cite{Pshirkov:2007st}, the finite electron density in the plasma within the NS magnetosphere gives the photon a mass $m_\gamma$.\footnote{Throughout this paper we work in natural units such that mass and frequency have the same dimensions; a plasma frequency of $\omega_p$ corresponds to a photon mass $m_\gamma = \hbar \omega_p/c^2$.} This is a key ingredient for obtaining resonant conversion, which can only take place when energy and momentum conservation are both satisfied. 
This photon mass is expected to fall off monotonically with distance from the NS surface. Thus for any axion mass smaller than the plasma mass at the surface, there exists a radius $r_c$, the conversion radius, at which the axion mass equals the plasma mass and the axion-photon conversion process takes place resonantly. However, the plasma around a NS is highly magnetized, with the cyclotron frequency $\Omega_c$ much larger than the plasma frequency $\omega_p$, and the effective photon mass is anisotropic with strong angular dependences. In this Letter, we solve the axion-photon equations of motion within the NS magnetosphere, including a full treatment of the magnetized plasma, to calculate the axion-photon conversion probability.  We find that if the axion mass equals the photon mass at $r_c$, then conversion takes place resonantly over a distance $L \sim \sqrt{r_c v_c/ m_a}$, where $v_c$ is the axion velocity at $r_c$.

The process of axion-photon conversion in neutron-star magnetospheres is related to the detection mechanism utilized by axion helioscopes, such as the CAST experiment, to search for relativistic axions produced inside of the Sun which propagate to Earth~\cite{Anastassopoulos:2017ftl,Arik:2013nya,Arik:2015cjv}.  Solar axions have a thermal spectrum, $E \sim$ keV,  and can convert to photons within CAST's static magnetic field ${\bf B}$. When the momentum transfer $q = m_a^2 / (2 E)$ is much less than the length of the $B$-field region $L$ ($q L \ll 1$), then conversion takes place with probability \mbox{$P_{a \gamma} \sim g_{a \gamma \gamma}^2 {\bf B}^2 L^2$}. 
However, for large $m_a$ the conversion rate drops rapidly because it is no longer possible to satisfy both energy and momentum conservation simultaneously; the axion and photon have different dispersion relations, which becomes more apparent for larger values of $m_a$ and longer distances $L$.  To circumvent this issue and maintain sensitivity to high-$m_a$ axions, CAST fills the $B$-field region with $^4$He and $^3$He to give the photon a mass so that the axion and photon have the same dispersion relation.  By varying the pressure of the gas, the plasma frequency can be adjusted to scan over a range of $m_a$ values.  With this technique, CAST has set some of the strongest limits on axion-like particles in the mass range $m_a \sim 10^{-4}$ -- $10^0$ eV~\cite{Arik:2013nya,Arik:2015cjv}.    

Most previous efforts to detect axion DM have focused on \emph{direct} detection; see \cite{Irastorza:2018dyq} for a detailed review. The majority of such experiments utilize the coupling of the axion to electromagnetic fields, though some experiments, such as CASPEr~\cite{Budker:2013hfa}, directly use the axion coupling to nucleons.  For example, ADMX utilizes a microwave cavity to induce DM axion-photon conversion in the presence of an external $B$-field.  ADMX has already constrained axion DM scenarios in a narrow mass range around $10^{-6}$ eV, and future runs of ADMX should expand the sensitivity to axion masses in the range $m_a \sim 10^{-6}$ -- $10^{-5}$ eV~\cite{Shokair:2014rna,Rosenberg:2015kxa}.  The HAYSTAC collaboration will try to push the reach to masses as high as $10^{-4}$ eV using similar technology~\cite{Brubaker:2016ktl,Kenany:2016tta,Brubaker:2017rna}, while MADMAX \cite{Majorovits:2016yvk} will probe a similar mass range with dielectric haloscopes \cite{TheMADMAXWorkingGroup:2016hpc}. A separate set of experiments, such as ABRACADABRA~\cite{Kahn:2016aff,Foster:2017hbq} and DM-Radio~\cite{Chaudhuri:2014dla,Silva-Feaver:2016qhh}, are working to test the axion DM hypothesis at lower masses, potentially down to $10^{-9}$ eV, by exploiting the coupling $a\, {\bf E} \cdot {\bf B}$ in the limit where the axion wavelength is much larger than the size of the experiment. 
Our work complements these approaches by providing an avenue for \emph{indirect} detection of QCD axion DM in the mass range $\sim 0.2-40~\mu {\rm eV}$ utilizing existing and planned radio telescopes. 

NSs have long been recognized as promising targets for axion searches due to their strong magnetic fields.  Previous efforts have focused on either photon-axion conversion leading to spectral distortions in the outgoing electromagnetic emission~\cite{Chelouche:2008ta} or the conversion of thermal axions, produced inside of the NS, into photons in the NS magnetosphere~\cite{Morris:1984iz}. However, neither of these processes require the axion to be DM, and also neither are sensitive enough to probe the QCD axion. In particular, thermal axions are ultrarelativistic and hence cannot undergo resonant conversion in the magnetosphere \cite{Raffelt:1987im}, but DM axions are only mildly relativistic and resonant conversion is obtained over a broad range of parameters. Our work builds upon this previous work by calculating the radio flux from axion DM conversion into photons within the magnetosphere. Assuming that the axion makes up all of the DM, we show that radio searches may be sensitive to QCD-axion-strength couplings $g_{a \gamma \gamma}$.\footnote{Our work is similar in spirit to that of~\cite{Pshirkov:2007st}, which also considered DM axion-photon conversion within the NS magnetosphere. However, we disagree with the details of many aspects of their calculations.}
  
{\it Neutron Star Magnetosphere. ---}
The magnetic field in the vicinity of the NS surface is thought to be well described by a dipole configuration, with an axis $\m$ that is misaligned from the rotation axis (which we take to be the $z$-axis) by an angle $\theta_m$.  Charged particles are stripped from the surface of the NS at the magnetic poles and accelerated along open field lines, producing the non-thermal, pulsed radio and gamma-ray emission seen from pulsars.  These regions near the magnetic poles are characterized by a high-density, boosted plasma.  On the other hand, the NS ``lobes'' consist of closed magnetic field lines and likely much more modest plasma densities.  We will take a simplistic approximation to the neutron-star magnetosphere, inspired by Goldreich and Julian (GJ)~\cite{1969ApJ...157..869G}. The GJ model gives the minimum plasma density necessary in the presence of the NS magnetic field, by finding a self-consistent solution to Maxwell's equations when particles on magnetic field lines corotate with the star. 

Though originally proposed for aligned NSs with $\theta_m = 0$, the GJ derivation applies equally well to misaligned NSs, and gives a charge density
\begin{align}
n_c & = \frac{2 \mathbf{\Omega} \cdot \mathbf{B}}{e}\frac{1}{1 - \Omega^2 r^2 \sin^2 \theta}  \,,
\label{n_e}
\end{align}
where $\Omega = 2\pi/P$ with $P$ the NS spin period, and $\theta$ is the polar angle with respect to the rotation axis.  We will take the charge density as a rough estimate of the electron number density: $n_e = |n_c|$.\footnote{When the charge density is positive, it could consist only of protons leading to a much smaller plasma mass.  If this were the case, some results may become stronger because low-mass axions would convert closer to the NS radius, where the magnetic field is higher.}
The plasma frequency is $\omega_p \approx \sqrt{4 \pi \alpha n_e / m_e}$, so that within the GJ model 
\es{omegaGJ}{
\omega_p \approx \left( 1.5 \times 10^{2} \, \, \text{GHz} \right) \sqrt{ \left( {B_z \over 10^{14} \, \, {\rm G}} \right) \left( {1 \, \, \text{sec} \over P} \right) } \,,
}
where
\es{B_z_mis}{
B_z = {B_0 \over 2} \left( {r_0 \over r} \right)^3 \big[3 \cos \theta \, {\bf \hat m} \cdot {\bf \hat r} - \cos \theta_m \big] 
}
is the component of the magnetic field along the ${\bf \hat z}$ direction.  Note that
\es{mdr}{
{\bf \hat m} \cdot {\bf \hat r} = \cos \theta_m \cos \theta + \sin \theta_m \sin \theta \cos(\Omega t)
}
depends on time due to the rotation of the NS.  In \eqref{omegaGJ} we have neglected the relativistic correction in the denominator of~\eqref{n_e}, which can be important for millisecond pulsars but is typically a percent-level correction for the pulsars with large $P$ that we will be concerned with in this analysis. In practice, the true plasma density is likely more complicated than the simple GJ model.  In particular, there could be non-trivial time dependence and boosts within the plasma.  However, the GJ model provides a straightforward starting point for this analysis, which we hope can be improved in future work with more realistic models for the NS magnetosphere. In this analysis, we focus only on the region of closed field lines where the plasma is expected to be nonrelativistic, leaving the complications of boosted plasma near the magnetic poles to future work.

As we show below, the axion-photon conversion occurs resonantly within the vicinity of the conversion radius $r_c$, defined to be the radius at which the plasma frequency equals the axion mass. Using the expressions above, we find that in the GJ model the time-dependent $r_c$ is given by
\es{rc}{
& r_c(\theta, \theta_m, t)  = 224 \, \, \text{km} \times \big|3 \cos \theta \, {\bf \hat m} \cdot {\bf \hat r} - \cos \theta_m \big|^{1/3} \times \\
 & \left( {r_0 \over 10 \, \, \text{km}} \right) \times \left[  {B_0 \over 10^{14} \, \, {\rm G} } \ { 1\, \, \text{sec} \over P} \left( {1 \, \, \text{GHz} \over m_a} \right)^2 \right]^{1/3} \,.
} 

{\it Conversion Probability from Mixing Equations. ---} Since the axion DM starts out non-relativistic far away from the NS and is accelerated to semi-relativistic velocities at radius $r_c$, we can approximate the axion trajectories as radial.
In the Supplementary Material (SM) we give a set of physical arguments that may be used to understand the parametric dependence of the axion-photon conversion probability. Here, we calculate the conversion probability by solving the coupled wave equations for the axion-photon system in the presence of the interaction term $-\frac{1}{4}g_{a \gamma \gamma} a {\bf E} \cdot {\bf B}$ in the Lagrangian, which leads to mixing between the axion $a$ and the component of the photon vector potential $A_\parallel$ that is transverse to the axion's motion but coplanar with the magnetic field. 

Following~\cite{Raffelt:1987im}, we assume radial plane wave solutions of the form $a(r,t) = i e^{i \omega t - i k r} \tilde a(r)$ and $A_\parallel(r,t) = e^{i \omega t - i k r} \tilde A_\parallel(r)$, where $k^2 = \omega^2 - m_a^2$.
As we will show, the resonant conversion takes place in a narrow enough region around $r_c$ that we may neglect the $r$ dependence of $\omega$.
Similarly, while the dispersion relation for $k$ holds for both the axion and the photon at $r_c$, the photon dispersion changes away from the conversion radius due to the continuously varying plasma mass.  We account for both of these effects in turn.  The analytic arguments presented below are supported by a full numerical analysis in the SM, where we also derive the equations of motion for the coupled axion-photon system in the plasma.

Near $r_c$, we may use the WKB approximation $|\tilde A_\parallel''(r)| \ll k |\tilde A'(r)|$ and $|\tilde a''(r)| \ll k |a'(r)|$.  The mixing equations reduce to the first-order ordinary differential equation
\es{ode}{
\left[ -i \frac{d}{dr} + \frac{1}{2 k}
{
 \left( 
\begin{array}{cc}
m_a^2 - \xi \, \omega_p^2   & \Delta_B \\
\Delta_B & 0
\end{array}
\right)}
 \right] 
\left(\begin{array}{c}  \tilde A_\parallel \\ \tilde a \end{array}\right)= 0 \,,
}
where
\es{}{
\xi = {\sin^2 \tilde \theta \over 1 - {\omega_p^2 \over \omega^2} \cos^2 \tilde \theta } \,, \quad \, \Delta_B = B g_{a \gamma \gamma} m_a {\xi \over \sin \tilde \theta} \,,  
}
$\omega =m_a \sqrt{1 + v_c^2}$, and $k =m_a v_c$.  Above, we have defined $\tilde \theta$ to be the angle between the propagation direction ${\bf \hat r}$ and the magnetic field ${\bf B}$.

For $r \gg r_c$, the axion-photon system no longer strongly mixes, but the amplitude of $A_\parallel$ modulates due to the varying plasma frequency of the medium. This effect is familiar from wave mechanics as it is exactly analogous to the increasing amplitude of ocean waves as they approach the shore (though in our analysis, we are considering the opposite case of waves leaving the shore).  
The net effect is a suppression of the outgoing electromagnetic wave by a factor of $\sqrt{v_c}$ asymptotically far away from the NS, namely $A_\parallel(\infty) \sim \sqrt{v_c} A_\parallel(r_c) $.

To calculate the energy flux in electromagnetic radiation asymptotically far from the NS, we may use the formalism of transition amplitudes by analogy to time-dependent perturbation theory in the Schr\"{o}dinger equation, working to first order in $\Delta_B$~\cite{Raffelt:1987im}. Taking initial conditions $\tilde{A}_\parallel(r_0) = 0$ and $\tilde{a}(r_0) = a_0$, and neglecting the modulation of the outgoing electromagnetic wave for now,~\eqref{ode} gives
\es{TProb_0}{
p_{a \gamma}(r) =&\left| \int_0^r dr' {B(r') \xi(r') g_{a \gamma \gamma} \over 2 v_c  \sin \tilde \theta}  \right. \\
&\left.\times e^{{-i \int_0^{r'} d\tilde r \big[m_a^2 - \xi(\tilde r) \omega_p^2(\tilde r) \big] \over 2 m_a v_c}  } \right|^2. 
}  
This expression represents $|A_\parallel(r)|^2/a_0^2$ and may be interpreted classically as the ratio of energy density in the electromagnetic field to the energy density in the axion field at a radius $r$.  Taking $r \to \infty$ and including the amplitude modulation of the outgoing electromagnetic field, we evaluate~\eqref{TProb_0} by the method of stationary phase to obtain
the energy transfer function 
\es{TProb}{
p_{a  \gamma}^\infty 
&\approx \frac{1}{2v_c} \, g_{a \gamma \gamma}^2 B(r_c)^2 L^2 \,,
}
with $L = \sqrt{2 \pi r_c v_c / (3 m_a)}$ and $p_{a\gamma}^\infty \equiv v_c \lim_{r\to\infty} p_{a\gamma}(r)$.  Note that $L$ may be interpreted as the distance over which the resonant conversion  takes place at $\tilde \theta = \pi / 2$.  While derived for $\tilde \theta = \pi/2$, the expression in~\eqref{TProb} holds for generic $\tilde \theta$ to leading order in $v_c$.

{\it Radiated Power. ---}
Next, we calculate the electromagnetic power emitted from the NS by axions converting into photons. We first note that since the NS plasma is optically thin, Thomson scattering of photons is negligible for the long-period NSs under consideration~\cite{Raffelt:1987im}, and thus outgoing photons do not scatter. Because $m_a^{-1} \ll r_c$ for $m_a$ in the MHz--GHz range, $L$ is parametrically smaller than $r_c$, and thus conversion takes place in a small region around $r_c$. We thus estimate the radiated power $\mathcal{P}$ by multiplying the flux of DM through a surface subtending a solid angle $d\Omega$ at $r_c$ by the energy transfer function:
\es{dPdOmegarc}{
{d\mathcal{P}(\theta,\theta_m t) \over d \Omega}  \approx 2 \times p_{a \gamma}^\infty \, \rho_\text{DM}^{r_c} v_c r_c^2 \,,
}
where $\rho_{\rm DM}^{r_c}$ is the DM mass density at $r = r_c$ and $v_c$ is the DM velocity at $r = r_c$. We note that all quantities on the right-hand side of \eqref{dPdOmegarc} depend on $\theta$, $\theta_m$, and $t$ through their dependence on $r_c$ (see~\eqref{rc}). The factor of two comes from the fact that the DM may convert into photons either on its way in to the NS or out of the NS; if it is converted on the way in, then the photon is reflected back out, since the higher-density plasma acts as a mirror to photons of frequency less than the plasma frequency.  

Let us assume that asymptotically far away from the NS, the DM has density $\rho_\text{DM}^\infty$ and is described by a Maxwell-Boltzmann velocity distribution with 
\es{f_infty}{
f_\infty( {\bf v_\infty}) = {1 \over \pi^{3/2} v_0^3} e^{- {\bf v_\infty}^2 \over v_0^2} \,,
}
where $v_0 \sim 10^{-3}$ is the DM virial velocity. We can calculate $v_c$ by conservation of energy: $v_c^2 = v_\infty^2 + {2 G M_{\rm NS} \over r_c} \approx {2 G M_{\rm NS} \over r_c}$, where $v_\infty$ is the DM speed asymptotically far away from the NS, which is typically much smaller than the escape velocity $\sqrt{2GM_{\rm NS}/r_c}$. Liouville's theorem maps the  phase-space distribution from asymptotic infinity to $r_c$: 
\es{LT}{
\rho^{r_c}_\text{DM} f_{r_c} ({\bf v}) = \rho_\text{DM}^\infty f_\infty ( {\bf v_\infty} [ {\bf v} ] ) \,,
}
where ${\bf v_\infty} [ {\bf v} ]$ denotes the velocity at asymptotic infinity that gives velocity ${\bf v}$ at radius $r_c$.  Integrating~\eqref{LT} and expanding in the limit $v_0^2 / (G M_{\rm NS} / r_c) \ll 1$ gives
\es{}{
\rho_\text{DM}^{r_c} = \rho_\text{DM}^\infty {2 \over \sqrt{\pi}} {1 \over v_0} \sqrt{2 G M_{\rm NS} \over r_c} + \cdots \,.
}
This then leads to an expression for the radiated power:
\es{power}{
&{d \mathcal{P} (\theta = {\pi \over 2}, \theta_m = 0)\over d \Omega} \approx  4.5 \times 10^{8} \, \, \text{W} \left( {g_{a \gamma \gamma} \over 10^{-12} \, \, \text{GeV}^{-1} } \right)^2  \\
&\left( {r_0 \over 10 \, \, \text{km} } \right)^2 \left( {m_a \over 1 \, \, \text{GHz} } \right)^{5/3} \left( {B_0 \over 10^{14} \, \, \text{G} } \right)^{2/3} \left( {P \over 1 \, \, \text{sec} } \right)^{4/3} \\
&  \left( {\rho_\infty \over 0.3 \, \, \text{GeV} / \text{cm}^3} \right) \left( {M_{\rm NS} \over 1 \, \, M_\odot} \right) \left( {200 \, \, \text{km}/\text{s} \over v_0} \right)\,,
}
with 
\es{power-theta}{
{d \mathcal{P}(\theta, \theta_m, t)\over d \Omega} = &{d \mathcal{P} (\theta ={ \pi \over 2}, \theta_m = 0) \over d \Omega}\\
 &\times  {3 \, ({\bf \hat m} \cdot {\bf \hat r})^2 + 1 \over \big|3 \cos \theta \, {\bf \hat m} \cdot {\bf \hat r} - \cos \theta_m \big|^{4/3} } \,.
}
Both~\eqref{power} and~\eqref{power-theta} are formally only valid so long as $r_c > r_0$; no resonant conversion takes place inside the NS.  This regulates the otherwise-divergent denominator in~\eqref{power-theta}, and also gives a strong angular and time dependence to the signal. 

{\it Radio Telescope Sensitivity. --- }
The radio flux at Earth is given by \mbox{$F(\theta, \theta_m, t) = d\mathcal{P}(\theta, \theta_m, t) / d\Omega / d^2$}, where $d$ is the distance from us to the NS.   However, the quantity that is more relevant for radio observations is the flux density, defined as $S = F / B$, where $B$ is the bandwidth.  Energy conservation implies that the expected bandwidth of the signal is $B \sim (v_0 / c)^2 m_a / (2 \pi)$.  This is because the frequency of the emitted photon asymptotically far away from the NS must be equal to the energy of the infalling DM particle that created it, regardless of the radius at which the DM was converted. In other words, the kinetic energy gained by the infalling DM is exactly canceled by the gravitational redshift of the outgoing photon, and the bandwidth of the signal is determined by the velocity dispersion in the \emph{asymptotic} DM distribution.  This leads to the estimate for the flux density 
\es{eq:Sfid}{
S &= 6.7 \times 10^{-5} \, \, \text{Jy} \, \left({100 \, \, \text{pc} \over d} \right)^2  \left( {1 \, \, \text{GHz} \over m_a} \right) \times
\\
& \left({ 200 \, \, \text{km}/\text{s} \over v_0} \right)^2\left[\frac{d\mathcal{P}/d\Omega}{4.5 \times 10^{8} \ {\rm W}}\right] \,.
}
This should be compared to the minimum detectable flux at a radio telescope, which is given by 
\es{}{
S_\text{min} = \text{SNR}_\text{min} { \text{SEFD} \over \sqrt{n_\text{pol} B \Delta t_\text{obs}} } \,,
}
where $ \text{SNR}_\text{min}$ is the minimum signal-to-noise ratio, $\text{SEFD}$ is the system-equivalent flux density, $n_\text{pol}$ is the number of polarizations (we will take $n_\text{pol} = 2$), $B$ is the bandwidth, and $\Delta t_\text{obs}$ is the observation time. As an example, the Arecibo Telescope has $\text{SEFD} \sim 2$ Jy.  Eq.~\eqref{eq:Sfid} holds for sources whose bandwidth is wider than the intrinsic frequency resolution of the telescope, which we will assume is always the case. In the SM we discuss the optimization of the bandwidth for axion DM radio signals and how to account for the non-trivial time-dependence of the light curve in our sensitivity estimates.

\begin{figure}[htb]
\includegraphics[width = 0.5\textwidth]{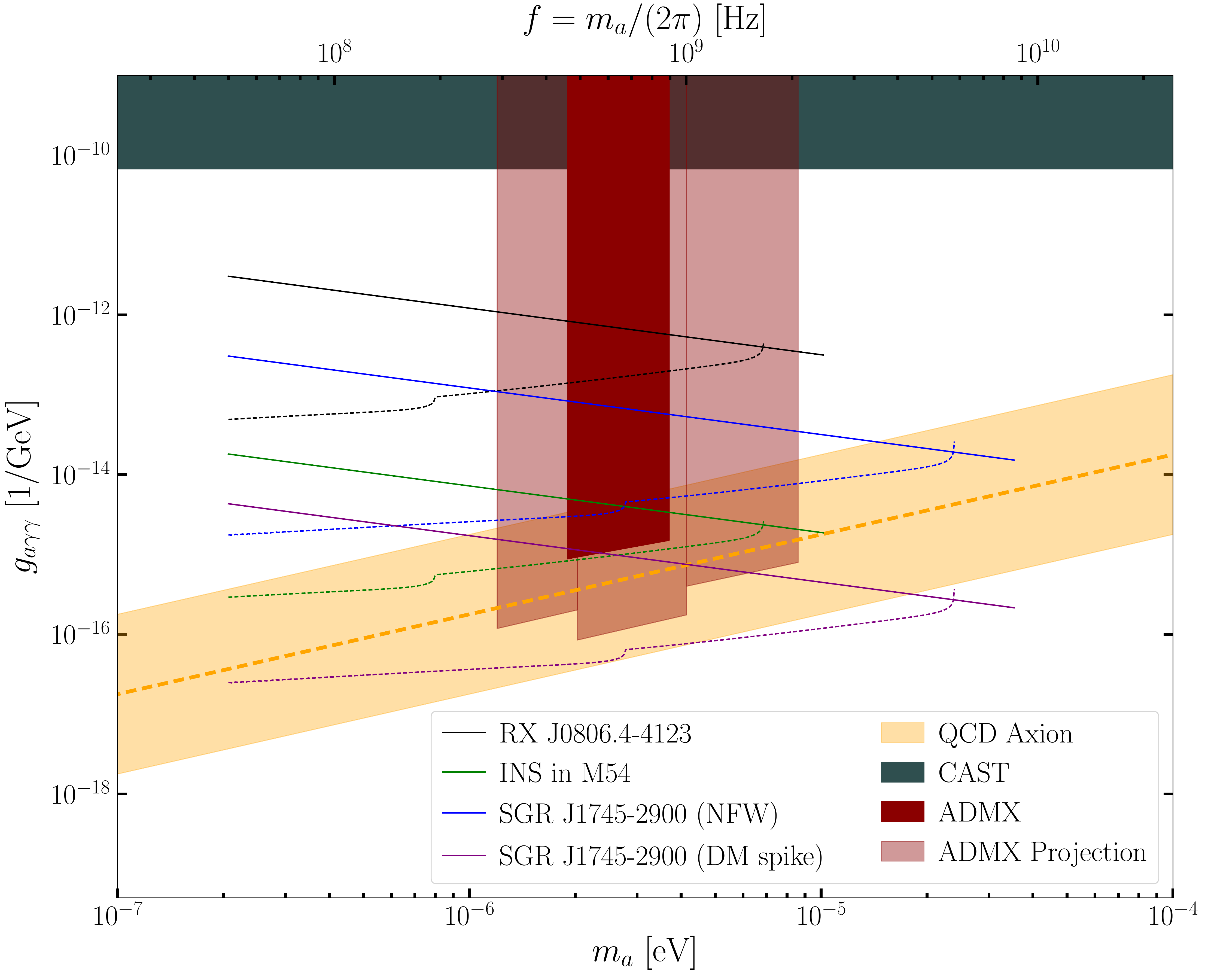}
 \vspace{-0.7cm}
\caption{The projected sensitivity to $g_{a \gamma \gamma}$ as a function of the axion mass $m_a$ for $\Delta t_\text{obs} = 100$ hr and $\text{SEFD} = 2$ Jy.  The QCD axion is predicted to lie within the band, while limits from CAST and ADMX (current and projected) are indicated.   
We have taken $\theta_m = 10^\circ$ and the solid (dashed) curves assume $\theta = 90^\circ$ ($\theta = 120.5^\circ$).  
The lower mass cutoff is set by the lowest available frequency of current radio telescopes, while the high-mass cutoff comes from requiring the conversion radius to be outside the NS radius.
}
\label{fig: limits}
\vspace{-0.5cm}
\end{figure}

{\it Neutron star targets. ---} As pointed out in~\cite{Pshirkov:2007st}, nearby isolated neutron stars (INS) make excellent targets for radio signals from axion DM conversion.  This group of $\sim$7 NSs is characterized by their proximity to Earth ($\lesssim 500$ pc), strong magnetic fields ($\sim$$10^{13}$ G), long spin periods ($\sim$5 s), and lack of observed pulsed, non-thermal emission (see, {\it e.g.},~\cite{1999A&A...349..389V,Kaplan:2008qn,Kaplan:2009ce}).  
Importantly, since these NSs do not exhibit radio emission, we can estimate the sensitivity to axion DM assuming that we are limited by thermal noise in the radio telescope rather than background radiation from the NS.  Additionally, the lack of non-thermal emission suggests that pair production at the NS surface is inefficient and that the GJ model for the plasma density may hold throughout the magnetosphere~\cite{Pshirkov:2007st}.

In Fig.~\ref{fig: limits} we show the sensitivity in $g_{a\gamma\gamma}$ 
from 100 hrs of observation of one of the isolated NSs, J0806.4-4123.  This NS has a period $P \approx 11.37$ s, magnetic field $B_0 \approx 2.5 \times 10^{13}$ G,
and is at a distance $d \approx 250$ pc from Earth~\cite{Kaplan:2009ce}. We also assume $M_{\rm NS} = 1 M_{\odot}$ and $r_0 = 10 \ {\rm km}$. 
 We take $\text{SEFD} \approx 2$ Jy for our estimates, 
 though this may be improved with future instruments such as the Square Kilometer Array.  Our sensitivity curves are defined by $1\sigma$ significance, as discussed in the SM. We show sensitivities calculated
for two pulsar geometries. The solid curve takes a generic value $\theta = 90^\circ$ for the polar angle of Earth in the NS frame, while the dashed curve is tuned to $\theta = 120.5^\circ$, which gives the near maximal signal at low masses and which is also highly pulsed from emission at orientations where $r_c \to r_0$. In both cases we take a generic misalignment angle $\theta_m = 10^\circ$. 
Note that the low-mass cutoff is set to the $m_a/(2\pi) = 50$~MHz threshold of typical radio telescopes, while the high-mass cutoff is determined by the maximum mass for which the conversion radius is outside the NS radius.

Another class of targets are NSs that occupy regions of high DM density and/or low velocity dispersion.  For example, consider the magnetar SGR J1745--2900, which is located $R \approx 0.1$ pc away from the Galactic Center~\cite{Kennea:2013dfa,Mori:2013yda,Shannon:2013hla,Eatough:2013nva}.  This magnetar has a magnetic field $B_0 \approx 1.6 \times 10^{14}$ G  
and a period $P \sim 3.76$ s~\cite{Kennea:2013dfa,Mori:2013yda}.  While the magnetar was first discovered in X-rays, a highly pulsed and variable $\sim$mJy radio signal has been observed from the magnetar (see, {\it e.g.},~\cite{Shannon:2013hla,Eatough:2013nva}).   The DM density in this region is highly uncertain.  Using the NFW and Burkert fits from~\cite{Nesti:2013uwa}, we find that the DM density at $R = 0.1$ pc is enhanced by a factor of $2 \times 10^5$ for the best-fit NFW profile, relative to the local density, but only a factor $\sim$4 for the best-fit cored Burkert profile. 
The cored profiles, however, may be in tension with new data from the Galactic bulge (see, for example,~\cite{2015MNRAS.448..713P,Hooper:2016ggc}).  On the other hand, if the DM distribution is described by a generalized NFW profile with an index $\gamma = 1.5$, which is allowed by the kinematic data available, then the enhancement would be $\sim$$10^7$.   
There is also the possibility of a DM density spike near Sgr A*, the supermassive black hole at the center of the Galaxy.  With the density spike, the DM density at $R = 0.1$ pc could be enhanced by a factor $\sim$$10^9$ 
relative to the local density~\cite{Lacroix:2018zmg}.

In Fig.~\ref{fig: limits}, we show the projected sensitivity from 100 hrs observation of SGR J1745-2900, assuming both the NFW DM profile (blue) and spike profile (purple), with solid and dashed lines representing the two geometries $\theta = 90^\circ$ and $\theta = 120.5^\circ$.  We take $v_0 = 200~{\rm km}/{\rm s}$ and $d = 8.5~{\rm kpc}$ for the distance to the Galactic Center and assume $M_{\rm NS} = 1 M_\odot$ and $r_0 = 10$~km as before. Despite the fact that pulsed radio emission has been observed from this magnetar, we have made these sensitivity estimates under the assumption that the dominant noise source is the thermal noise in the telescope. Since the non-thermal radio emission is pulsed, non-pulsed (or pulsed but out-of-phase) DM-induced flux would likely still be dominated by telescope noise.  Interestingly, as seen in Fig.~\ref{fig: limits}, observations of SGR J1745-2900 could be sensitive to the QCD axion over multiple orders of magnitude in $m_a$, depending on the DM density profile.  However, we stress that this sensitivity estimate relies on the GJ model, which may not apply to the magnetar.

Alternatively, one could consider isolated NSs within dwarf galaxies. In the Sagittarius dwarf galaxy, the central DM density is enhanced by a factor $\sim$$5 \times 10^{5}$ compared to the local density in the solar neighborhood, 
and the velocity dispersion is low, $v_0 \sim 10$~km/s \cite{Aharonian:2007km}. For this estimate we have taken the cored DM density profile from~\cite{Aharonian:2007km}.  The globular cluster M54 appears to be coincident with the center of the Sagittarius dwarf galaxy, with the cluster having a core radius $\sim$1 pc, a mass $\sim$$2 \times 10^6$ $M_\odot$, and a distance of around $\sim$20 kpc from Earth~\cite{Monaco:2004ke,2009ApJ...699L.169I}.  
Given the mass of M54, there are likely many hundreds of NSs within the central core~\cite{Ivanova:2007bu}.  Assuming that just one of these NSs has the properties of J0806.4-4123, we would obtain the sensitivity to $g_{a \gamma \gamma}$ shown in Fig.~\ref{fig: limits} (labeled INS in M54).  If there are $N$ such INS's in the field of view, then we may expect the sensitivity to improve as $1/\sqrt{N}$.  The fact that all NSs radiate at the same frequency from axion DM could make even more distant galaxies promising targets.

A narrow radio line from a NS target could provide a striking signature of axion DM.  On the other hand,
in the absence of a signal, it will be difficult to set a robust limit on $g_{a \gamma \gamma}$ because of challenges in understanding confidently the plasma density and time-dependent dynamics in the inner regions of the magnetosphere.  Towards that end, it would be useful to incorporate the physics of axion-photon conversion into NS simulations~\cite{Philippov:2013tpa}.  Such work should lead to more precise predictions for the radio-line signal.

{\it Note added ---} While this work was in the final stages of preparation, Ref.~\cite{Huang:2018lxq} appeared, which addresses similar questions. Our work differs in several respects, but importantly where we do overlap we disagree in detail with their results for the conversion probability, radio flux, and projected sensitivity. 

{\it Acknowledgments. ---} YK and BS thank Jesse Thaler for collaboration in the early stages of this project. We thank Anatoly Spitkovsky for detailed discussions regarding NS magnetospheres, and Nahum Arav, Kfir Blum, Junwu Huang, Paul Ray, Nicholas Rodd, Jonathan Squire, Christoph Weniger, and Kathryn Zurek for useful discussions. 

\bibliography{Axion}

\clearpage
\newpage
\maketitle
\onecolumngrid
\begin{center}
\textbf{\large Radio Signals from Axion Dark Matter Conversion \\
in Neutron Star Magnetospheres} \\ 
\vspace{0.05in}
{ \it \large Supplementary Material}\\ 
\vspace{0.05in}
{}
{ Anson Hook, Yonatan Kahn, Benjamin R. Safdi, Zhiquan Sun}

\end{center}
\setcounter{equation}{0}
\setcounter{figure}{0}
\setcounter{table}{0}
\setcounter{section}{1}
\renewcommand{\theequation}{S\arabic{equation}}
\renewcommand{\thefigure}{S\arabic{figure}}
\renewcommand{\thetable}{S\arabic{table}}
\newcommand\ptwiddle[1]{\mathord{\mathop{#1}\limits^{\scriptscriptstyle(\sim)}}}

This Supplementary Material contains additional calculations and examples that support the results presented in the main Letter and further illustrate the phenomenology of a hypothetical signal. We begin by giving a heuristic derivation of the conversion probability that does not rely on having to solve the coupled differential equations but rather is based on more general physics arguments.  We also give an extended derivation of the mixing equations and their solution, including the effects of strong magnetic fields in the plasma, along with numerical examples that support the analytic solution for the conversion probability used in the main Letter. Next, we discuss the  non-trivial light curves and polarization profiles that might be expected from an axion signal.  Finally, we describe how to estimate the radio sensitivity, accounting for the optimal bandwidth for an analysis of the radio data given knowledge of the DM velocity distribution and also knowledge of the non-trivial light curves, and highlight the strong angular dependence of the signal.

\section{Conversion probability parametrics}

We first demonstrate how to understand the parametric dependence of the axion-photon conversion rate.  Parametrically, the axion photon conversion rate is of the form
\es{CAST_P}{
P(a \rightarrow \gamma) \sim \sin^2 \Theta \sin^2 \left ( \Delta k L \right )
}
where $\tan \Theta \sim B g_{a \gamma \gamma} \omega/(m_a^2-m_\gamma^2)$ is the mixing between the axion and the photon, $\Delta k$ is the difference in momentum between an axion and photon of the same energy, and $L$ is the distance over which the conversion occurs.  As is well known, relativistic axions converting in a magnetic field in vacuum have $m_\gamma = 0$ and $\Delta k \sim m_a^2/\omega$ so that $\Theta \ll 1$ and the conversion probability scales as 
\es{Eq: CAST}{
P(a \rightarrow \gamma) \sim \Theta^2 \Delta k^2 L^2 \sim B^2 g_{a \gamma \gamma}^2 L^2 .
}
This gives the conversion probability familiar from experiments such as CAST.

In this work, we consider the very different regime where axion-photon mixing is maximal due to $m_\gamma \sim m_a$, but the axion is non-relativistic.
As a direct consequence of the assumption that the photon effective mass is the same as the axion mass, we have $\Theta \sim \mathcal{O}(1)$.  When mixing is maximal, the difference in momentum between the two propagating eigenstates is $\Delta k \sim \frac{B \omega g_{a \gamma \gamma}}{k}$, where we are using~\eqref{ode} (or~\eqref{ode2} below) and ignoring any $\tilde \theta$ dependence for simplicity.  Thus the conversion rate is given by
\es{Eq: NS}{
P(a \rightarrow \gamma) \sim \Delta k^2 L^2 \sim \frac{1}{v_c^2} B^2 g_{a \gamma \gamma}^2  L^2.
}
Note that while~\eqref{Eq: NS} is similar to~\eqref{Eq: CAST}, the derivation and region of validity is completely different.  For example, if the axions were assumed to be relativistic instead of non-relativistic, the conversion probability would instead be highly suppressed by the large magnetic fields of the NS~\cite{Raffelt:1987im}.   Because NSs are macroscopic objects with large magnetic fields that extend over macroscopic distances, the length scale $L$ is not determined by the change in the magnetic field, but instead by the change in the plasma frequency of the photon.  Eq.~\eqref{Eq: NS} assumes coherent conversion, i.e.\ that photons which were generated from conversion at $r$ and $r+L$ add coherently.  If the plasma mass changes over this distance, then photons generated at $r_1$ can instead interfere with photons generated at a different radius $r_2$.  $L$ is roughly the distance over which photons generated at the beginning start to interfere with photons generated at the end,  i.e.\ $L \sim 1/\delta k $, where $\delta k$ is the difference in momentum at different locations.  $k \delta k \sim m_\gamma \delta m_\gamma$ and due to the power-law dependence of the photon effective mass on $r$, we also have $\delta m_\gamma/m_\gamma \sim L/r_c$.  Thus we have the estimate that
\es{our_L}{
L \sim \sqrt{\frac{r_c v_c}{m_a}}.
}
We note that our result is parametrically different from the estimate in \cite{Pshirkov:2007st}, which assumes $L \sim r_c$.
The final axion photon conversion rate is
\es{us}{
P(a \rightarrow \gamma) \sim \frac{B^2 g_{a \gamma \gamma}^2 r_c}{v_c m_a}.
}
In the main Letter, and in more detail below, we carefully calculate the axion-photon conversion probability from the mixing equations, but the parametric scaling can be understood by the arguments just put forth.

\section{Axion-photon conversion}
In this section, we give additional details of the equations of motion of the axion-photon system and the approximate solution to these equations for axion DM in the NS magnetosphere. 

\subsection{Equations of motion}
The plasma surrounding a NS is a cold, highly-magnetized plasma and due to the large magnetic field it is strongly birefringent.  
The effect of the magnetic field and plasma is taken into account by introducing a dielectric tensor.  Because the photon has both
transverse and longitudinal modes, it is convenient to work directly with the electric field.  The propagation of electromagnetic waves and axions in a plasma is determined by
\es{pde}{
-\partial_t^2 a + \nabla^2 a = m_a^2 a - g_{a \gamma \gamma} \mathbf{E} \cdot \mathbf{B}, \\
- \nabla^2 \mathbf{E} + \nabla ( \nabla \cdot \mathbf{E} ) = \omega^2 \mathbf{D} + \omega^2 g_{a \gamma \gamma} a \mathbf{B},
}
where the magnetic field $\mathbf{B}$ is taken to be the external magnetic field due to the NS.  We work over a small enough region of space where the change in gravitational potential may be neglected.
The electric displacement field $\mathbf{D}$ is given by \cite{gurevich2006physics}
\es{}{
\mathbf{D} = R^{yz}_{\tilde \theta} \cdot \begin{pmatrix}
    \epsilon      & i g & 0  \\
    - i g & \epsilon & 0  \\
    0 & 0 & \eta
\end{pmatrix} \cdot R^{yz}_{- \tilde \theta} \cdot \begin{pmatrix}
    E_x \\
    E_y   \\
    E_z 
\end{pmatrix},
}
where the magnetic field is taken to be at an angle $\tilde{\theta}$ from the $z$-axis in the positive $y-z$ quadrant, and $R^{yz}_{\tilde \theta}$ is the rotation matrix by $\tilde \theta$ in the $yz$-plane. The coefficients in the dielectric tensor are given by
\es{}{
\epsilon = 1 - \frac{\omega_p^2}{\omega^2 - \Omega_c^2} \qquad g = \frac{\omega_p^2 \Omega_c}{\omega (\omega^2 - \Omega_c^2)} \qquad \eta = 1 - \frac{\omega_p^2}{\omega^2} \qquad
\omega_p = \frac{4 \pi \alpha n_e}{m_e} \qquad \Omega_c = \frac{\sqrt{\alpha} B}{m_e} \,.
}
Taking the Fourier transform in time, and going to the high-magnetization limit ($\Omega_c \gg \omega, \omega_p$), which sends $\epsilon \to 1$ and $g \to 0$, $E_x$ decouples from the equations, and $E_z$ does not propagate and can be solved for algebraically.  The mixing matrix simplifies to
\es{ode2}{
-\partial_z^2 \begin{pmatrix}
    E_y   \\
    a 
\end{pmatrix} =  \begin{pmatrix}
  \frac{\omega^2 - \omega_p^2}{1- \frac{\omega_p^2}{\omega^2}\cos^2 \tilde{\theta}} & \frac{\g B_t \omega^2 }{1- \frac{\omega_p^2}{\omega^2}\cos^2 \tilde{\theta}}  \\
  \frac{\g B_t }{1- \frac{\omega_p^2}{\omega^2}\cos^2 \tilde{\theta}}  & \omega^2 - m_a^2
\end{pmatrix} \cdot \begin{pmatrix}
    E_y   \\
    a 
\end{pmatrix},
}
where $B_t = |\mathbf{B}|\sin \tilde{\theta}$ is the component of the $B$-field transverse to the direction of motion. Equation~\eqref{ode} is derived from a WKB approximation of this second-order differential equation, assuming outgoing plane waves in the radial direction with a local coordinate system defined by $\hat{\bf{r}} = \hat{\bf{z}}$ and defining $A_\parallel = E_y/i \omega$.

Note that in \eqref{ode2}, we have neglected the vacuum birefringence term from strong-field QED. Ref.~\cite{Raffelt:1987im} points out that this term can drastically affect the axion-photon conversion probability; however, this term is negligible for non-relativistic DM axions. Vacuum birefringence changes the condition for resonance from $\omega_p = m_a$ to 
\be
\omega_p^2 - \frac{7}{2}\omega^2 \kappa \sin^2 \tilde \theta = m_a^2,
\label{eq:VacuumBiref}
\ee
where
\be
\kappa = \frac{\alpha}{45\pi} \left(\frac{B}{B_{\rm crit}}\right)^2
\ee
determines the strength of vacuum birefringence effects. The critical field is
\be
B_{\rm crit} = \frac{m_e^2}{e} \approx 4.4 \times 10^{13} \ {\rm G}.
\ee
Ref.~\cite{Raffelt:1987im} points out that for sufficiently large $\kappa$ and ultrarelativistic axions with $\omega \gg m_a$, the maximal mixing condition \eqref{eq:VacuumBiref} can never be satisfied because the left-hand side becomes negative. Fortunately, for nonrelativistic or mildly relativistic axions with $\omega = \mathcal{O}(1) \times m_a$, this is never a problem. Indeed, in this case we have
\be
\frac{7}{2}\omega^2 \kappa \sin^2 \tilde \theta  \sim m_a^2 \times 10^{-4} \left(\frac{B}{4.4 \times 10^{13} \ {\rm G}}\right)^2.
\ee
Even for the largest NS magnetic fields considered in the Letter, the effects of the vacuum birefringence term are a percent-level perturbation and can be neglected for semi-relativistic axions, and $\omega_p = m_a$ can always be satisfied. More specifically, as long as $B \lesssim 10^{15} \ {\rm G}$ (the size of the largest known magnetar fields), the maximal mixing conditions can always be satisfied, but for these large fields one would need to consider the effect of $\kappa$ when solving for $r_c$.

\subsection{Damping of the photon wave}

As mentioned in the main Letter, it is important to take into account the damping of the outgoing electromagnetic wave in calculating the energy transfer from the axion field to the outgoing electromagnetic radiation. 
To isolate this effect, we neglect the mixing terms and consider the photon wave equation with a spatially-dependent plasma mass:
\es{}{
 \frac{d^2}{d r^2} A_\parallel + \left( \omega^2 - \omega_p^2(r)\right) A_\parallel = 0 \,.
} 
We now define $k(r) = \sqrt{\omega^2 - \omega_p^2(r)}$ and use a second WKB approximation, performing an expansion in the small dimensionless parameter $\varepsilon = \frac{1}{k^2}\frac{dk}{dr}$. We have verified that this parameter is small for 
$r>r_c$. 
This gives the solution (for $r > r_c$)
\es{A_k}{
A_\parallel(r,t)  = \frac{c}{\sqrt{k(r)}} e^{i \omega t - i \int_{r_c}^r k(r') dr'}.
}
Solving~\eqref{ode} for the photon field leads to the expression in~\eqref{TProb}. 

\subsection{The axion-photon evolution at finite $r$}

In the main Letter, we presented an approximate expression for $p_{a \gamma}$ valid at $r \to \infty$.  In this subsection, we give the result at finite $r$.  First, we consider $\tilde \theta = \pi/2$, where $\tilde \theta$ is the angle between the $B$-field and ${\bf \hat r}$, and then we discuss the generalization to arbitrary $\tilde \theta$.  
Taking $\tilde \theta = \pi/2$ and performing the integral in~\eqref{TProb} at finite $r$, we find
\es{TProb_finite_r}{
p_{a  \gamma}(r) 
&\approx \frac{1}{2v_c^2} \, g_{a \gamma \gamma}^2 B(r_c)^2 L^2 \times G\left(\frac{r-r_c}{L}\right) \\
&\times \begin{cases} 1, & r < r_c  \\ 
\frac{m_a v_c}{k(r)}, & r \geq r_c \end{cases} \,,
}
with $L = \sqrt{2 \pi r_c v_c / (3 m_a)}$ and the function $G(x)$ defined by 
\es{g}{
G(x) ={\left(\frac{1}{2}+C(x)\right)^2 + \left(\frac{1}{2}+S(x)\right)^2 \over 2}
}
in terms of the Fresnel $C$ and $S$ integrals. 
Note that $\lim_{x \to \infty} G(x) = 1$ and that $G\left(\frac{r-r_c}{L}\right)$ rises quickly from $0$ at $r < r_c$ to values near unity at $r \sim r_c$, over a distance of order $L$, while the amplitude modulation of the outgoing wave is encapsulated by the second line in~\eqref{TProb_finite_r}.  

At arbitrary $\tilde \theta$, the length $L$, which is found from the stationary phase approximation to the integral in \eqref{TProb_0}, is modified to
\es{eq:Lthetatilde}{
L'(\tilde \theta) = L\left(\tilde \theta = {\pi \over 2}, r'_c \right) \times {\sin \tilde \theta \sqrt{ 1 + v_c^2} \over \sqrt{ 1 + v_c^2  \sin^2 \tilde \theta}} \,,
}
and we should also make the substitution in~\eqref{TProb_finite_r}
\es{eq:Bthetatilde}{
B'(\tilde \theta,r_c) = B(\tilde \theta = {\pi \over 2},r'_c) \times {1 + v_c^2 \sin^2 \tilde \theta \over \sin \tilde \theta (1 + v_c^2)} \,.
}
Moreover, the conversion radius $r_c$ is modified to
\es{}{
r'_c = r_c \times \left( {1 + v_c^2 \sin^2 \tilde \theta \over 1 + v_c^2 } \right)^{1/3} \,.
}
Note that $B(r_c)^2 L^2$, the quantity which appears in $p_{a\gamma}$,  has no explicit dependence on $\tilde \theta$ up to $\mathcal{O}(v_c^2)$ corrections.

\begin{figure}[t!]
\includegraphics[width = 0.45\textwidth]{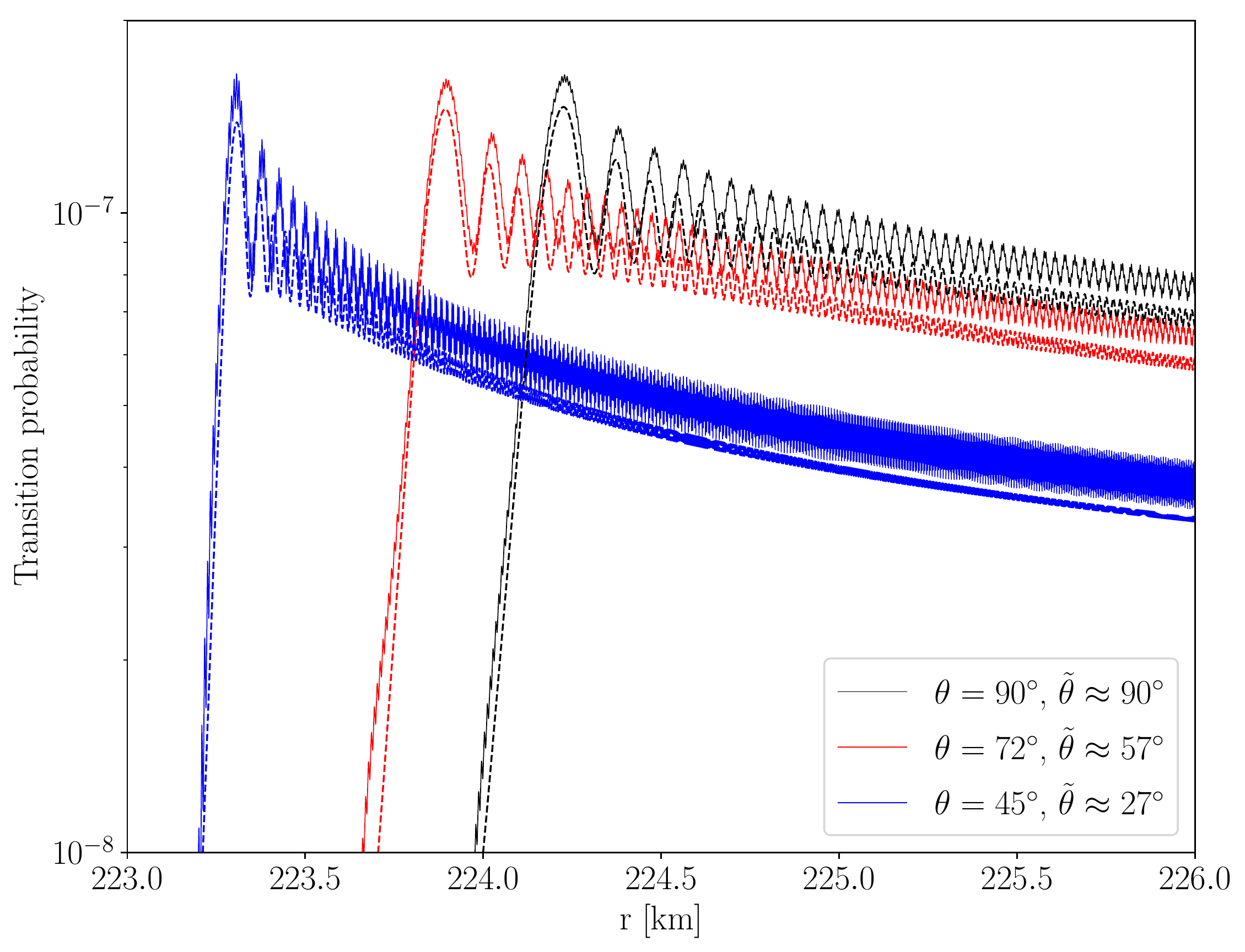} 
\includegraphics[width = 0.45\textwidth]{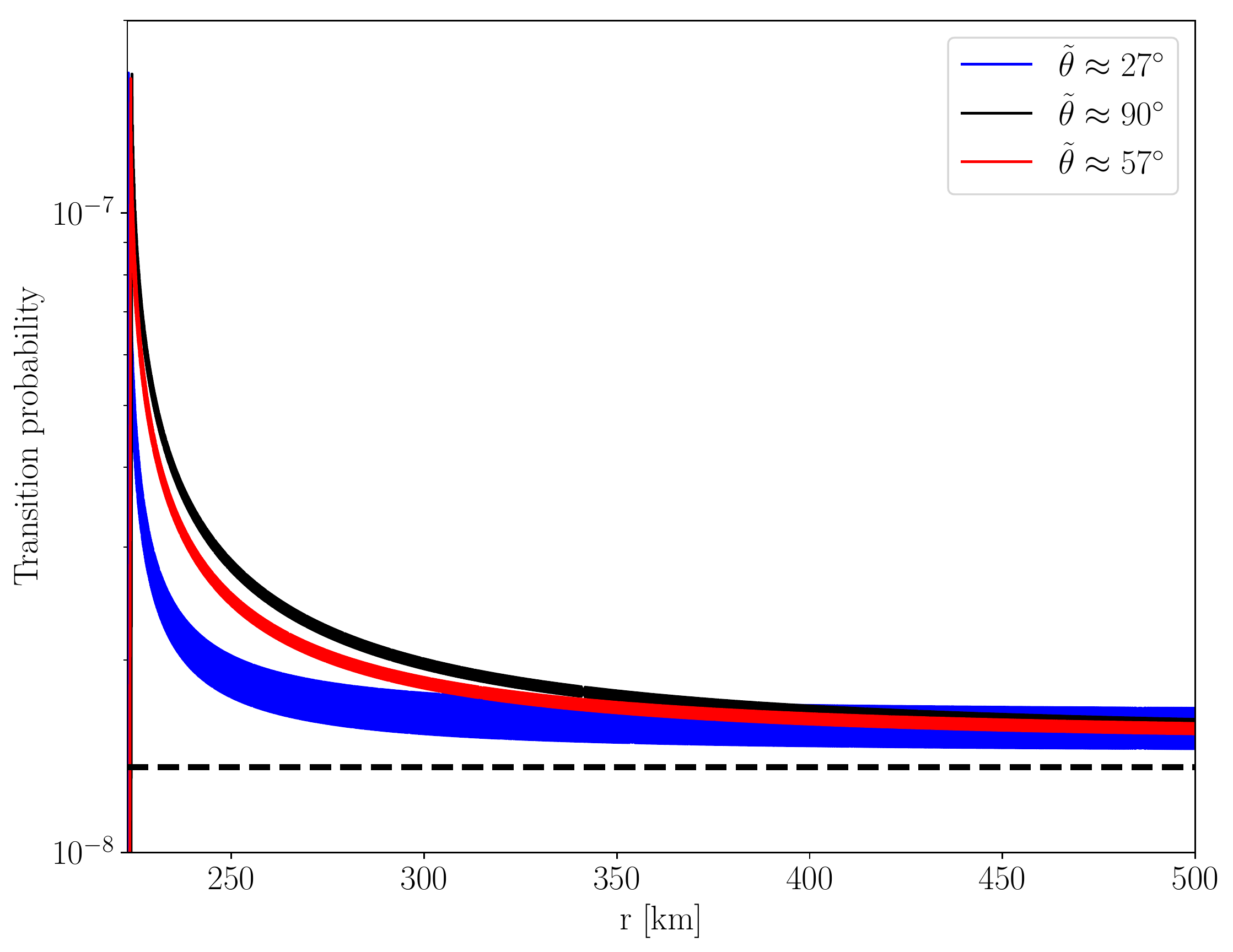}
 \vspace{-0.3cm}
\caption{Energy transfer function, also interpreted as a transition probability, from axions to photons as a function of the radius $r$ from the NS center, for parameters $r_0 = 10$ km, $M_\text{NS} = 1 \, M_\odot$, $B_0 = 10^{14}$ G, $\theta_m = 0$, $P = 1$ sec, $m_a = 1$ GHz, and $g_{a \gamma \gamma} = 10^{-12}$ GeV$^{-1}$.  Three different angles $\theta$ are shown, along with their corresponding angles $\tilde \theta$ between ${\bf \hat r}$ and the magnetic field.  {\it (Left)} The solid curves show the numerical solutions while the dashed illustrate the analytic approximation, for $r$ near the conversion radius.  {\it (Right)} As in the left panel, except over a wider range of $r$ and only illustrating the numerical solutions (solid) and the asymptotic approximation $p_{a \gamma}^\infty$ (black, dashed) at $r \to \infty$, which is the same for all $\tilde \theta$ to leading order in $v_c$.
}
\label{fig:solL}
\vspace{-0.2cm}
\end{figure}

In Fig.~\ref{fig:solL}, we compare our analytic estimate \eqref{TProb_finite_r} to numerical solutions to the full 2$^\text{nd}$-order mixing equations in~\eqref{ode2}, with boundary conditions imposed at $r = 223.05 \ {\rm km}$ ($r = 223.5 \ {\rm km}$) for the blue (black and red) curves. We take the frequency $\omega \approx m_a(1+ \frac{1}{2}v_c^2)$, where $v_c$ is the DM velocity at the conversion radius.
In this example, we take $r_0 = 10$ km, NS mass $M_{\rm NS} = 1 \, M_\odot$, $B_0 = 10^{14}$ G, $\theta_m = 0$, $P = 1$ sec, $m_a = 1$ GHz, and $g_{a \gamma \gamma} = 10^{-12}$ GeV$^{-1}$.  This implies that the transition radius is at $r_c \approx 224$ km and that at this radius the axion velocity is $v_c \approx 0.11$, in natural units, where we neglect the $\mathcal{O}(10^{-3})$ initial axion velocity $v_0$ asymptotically far away from the NS.  We illustrate three different angles, $\theta = 90^\circ$, $72^\circ$, and $45^\circ$; for each angle, the angle $\tilde \theta$ between ${\bf \hat r}$ and the magnetic field is indicated in the figure.  As we change $\theta$, we change the conversion radius $r_c$ given in~\eqref{rc} because of the angular dependence of the plasma frequency.  However, in order to highlight the differences between the different $\tilde \theta$ we chose to fix the normalization of the plasma frequency to that found at $\theta = 90^\circ$.  That is, at the level of the equations of motion in~\eqref{ode2}, we keep the implicit dependence of $\omega_p$ on $\theta$ fixed and only vary the explicit dependence of $\tilde \theta$.    In the left panel, the solid curves show the numerical solutions, while the dashed show the analytic approximations, which are seen to agree well. Indeed, we see that both the peak values and the asymptotic values for both the analytical and numerical solutions are independent of $\tilde \theta$ (and hence $\theta$).  This is further illustrated in the right panel, where we compare the numerical solutions to the asymptotic value for $p_{a \gamma}^\infty$ (black, dashed), which is the same for all $\tilde \theta$ to leading order in $v_c$. 

  We note that there are small discrepancies between the analytic approximation~\eqref{TProb} and the numerical solutions in these examples.  These may be due in part to the necessity of setting the boundary conditions very close to $r_c$ rather than at $r_0$ to avoid contamination of the solution by the spurious exponentially-growing mode when $\omega < \omega_p$.  Still, the difference between the analytic and numerical results is less than $\sim 10\%$ at large $r$.

\section{Light curves and polarizations}

\begin{figure}[t!]
\includegraphics[width = 0.45\textwidth]{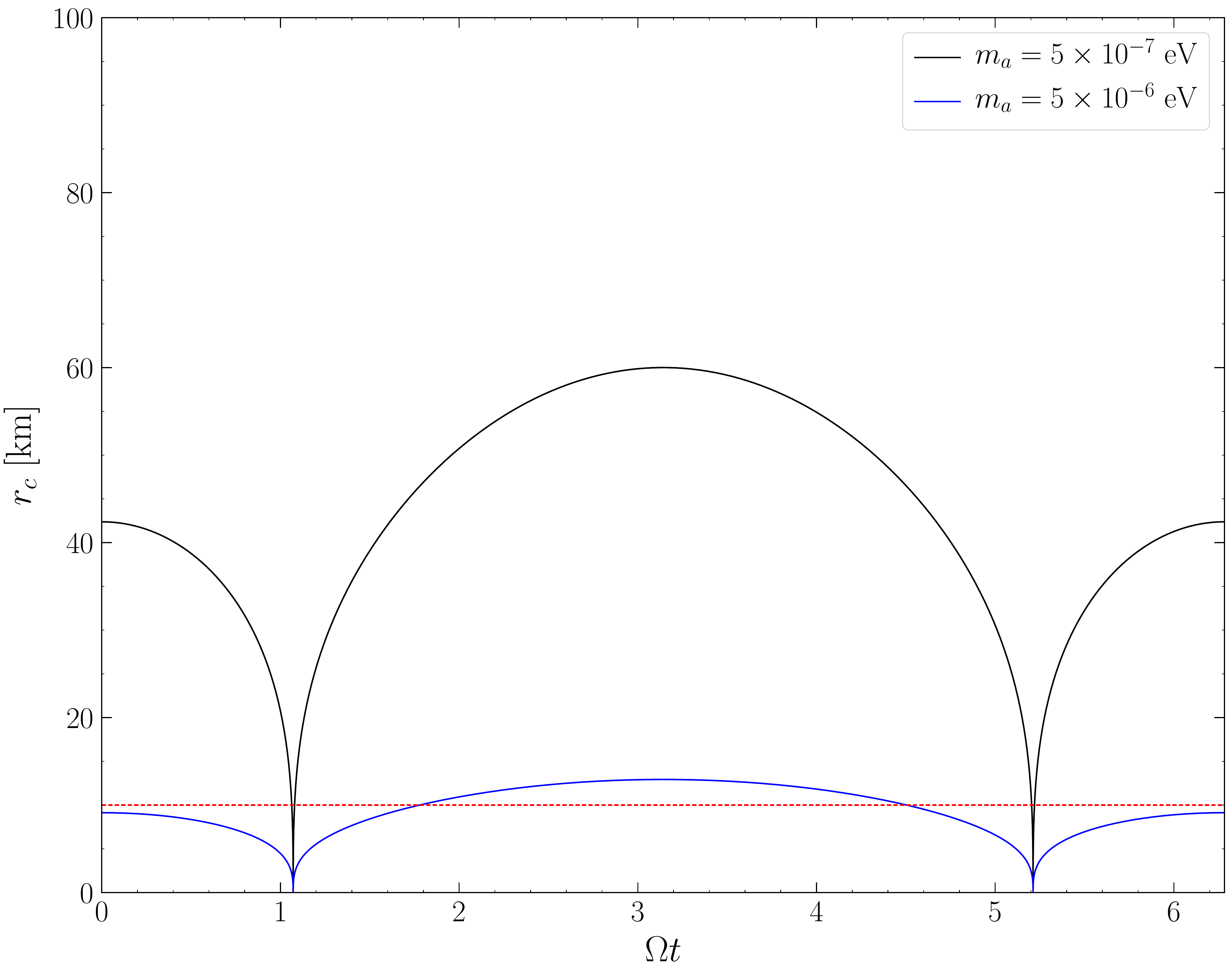} \includegraphics[width = 0.45\textwidth]{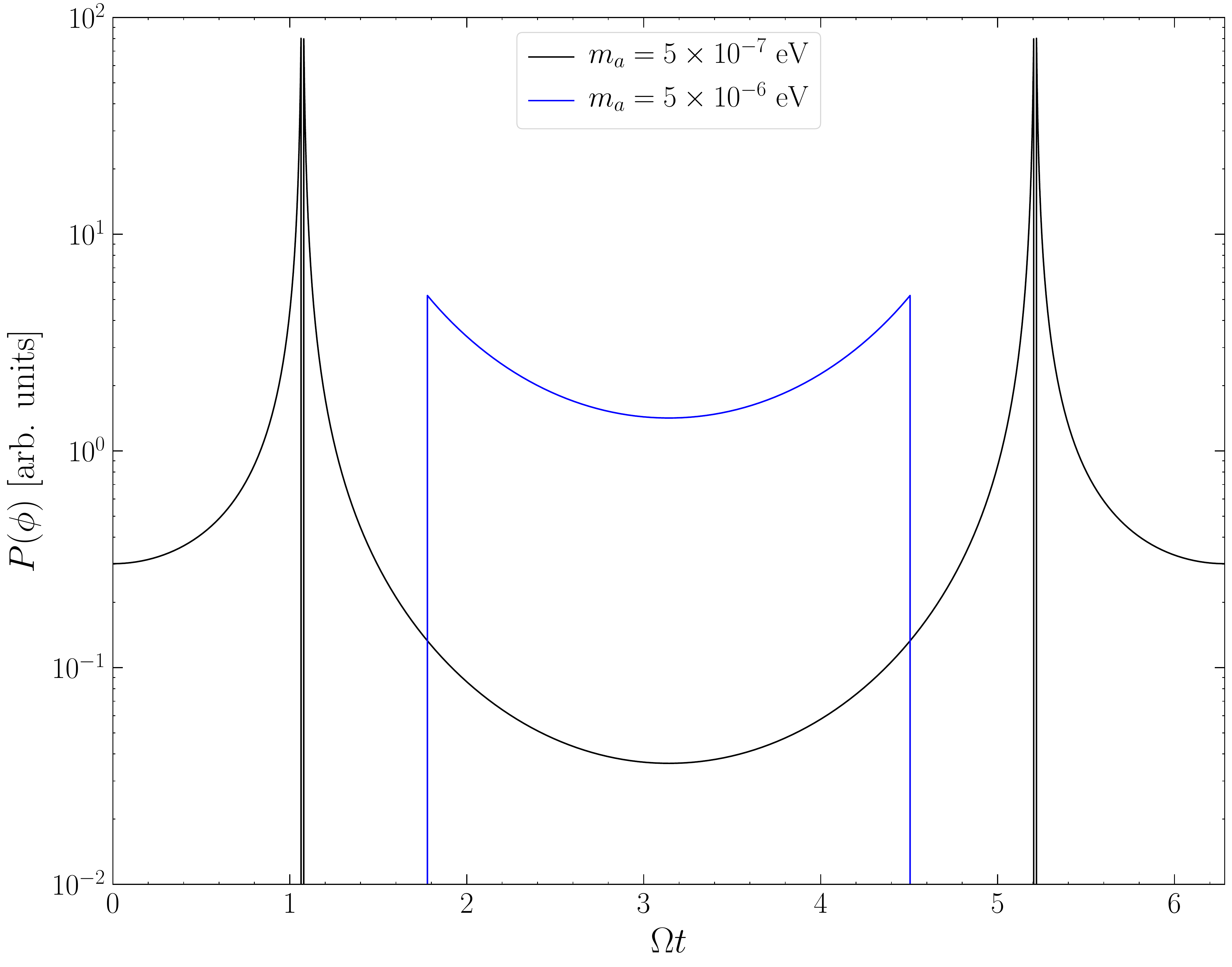}
 \vspace{-0.2cm}
\caption{(\emph{Left}) The conversion radius as a function of the phase $\Omega t$ for RX J0806.4-4123 assuming $\theta_m = 15^\circ$ and $\theta = 45^\circ$, for two different axion masses.  (\emph{Right}) The radiated power for the setup in the left panel, in arbitrary units, as a function of the phase.  The radiation stops when the conversion radius falls below the radius of the NS, which is indicated by the dashed red line in the left panel. 
}
\label{fig: conversion}
\end{figure}

The misalignment between the pulsar's magnetic axis and rotation axis leads to non-trivial light curves. Fig.~\ref{fig: conversion} shows the change in the conversion radius and the radiated power over the pulsar's rotation period as a function of the rotation phase for an example NS.

\begin{figure}[htb]
\includegraphics[width = 0.5\textwidth]{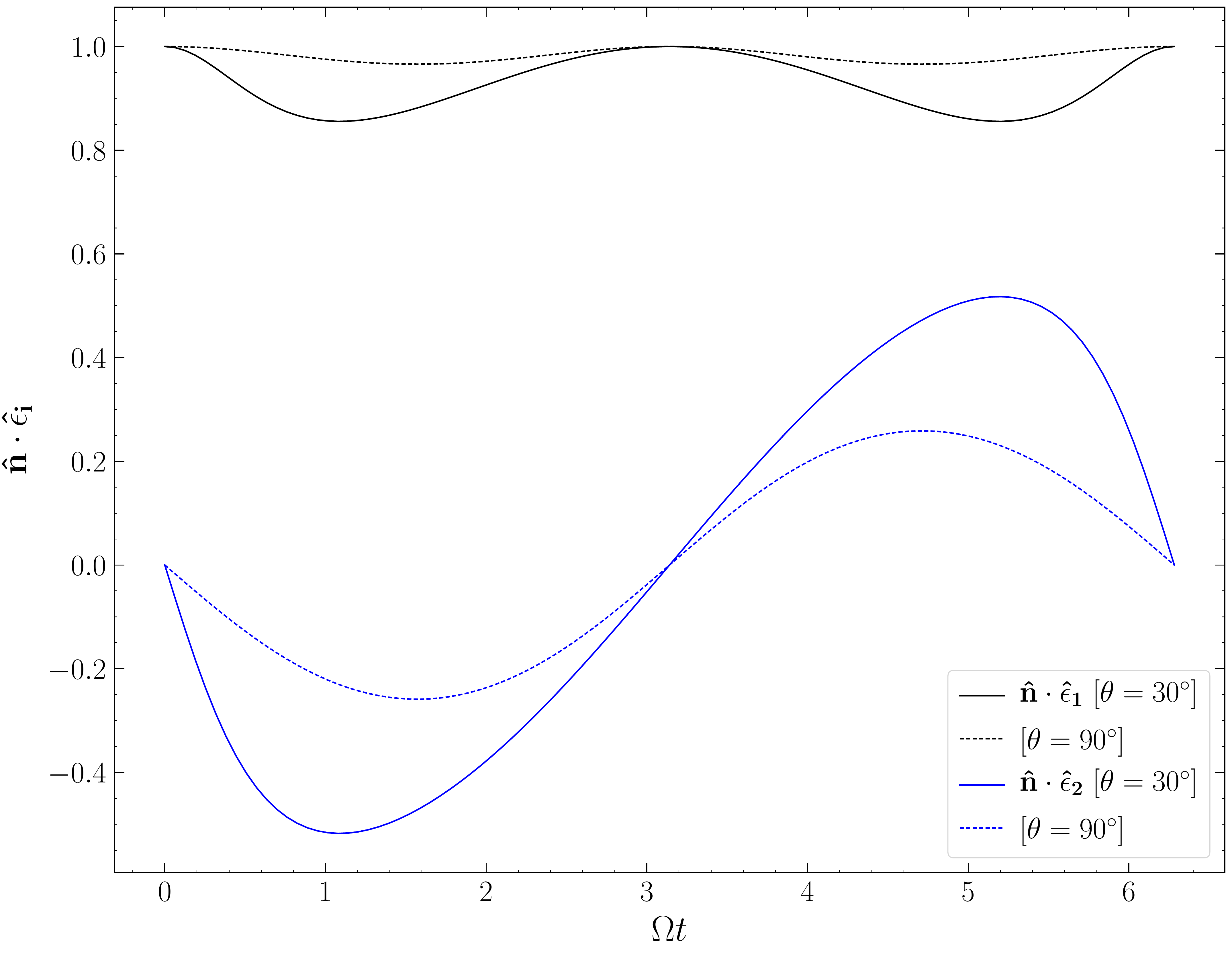}
\caption{The $\hat{\epsilon}_1$ (black) and $\hat{\epsilon}_2$ (blue) polarization components of the electric field as a function of the phase $\Omega t$ for various $\theta$, with $\theta_m = 15^\circ$. Note that these polarization curves are valid for any NS with these alignments, though depending on $m_a$ and the NS properties, the NS may not radiate over the whole period since $r_c$ may drop below the conversion radius (see Fig.~\ref{fig: conversion}).
}
\label{fig: polarization}
\end{figure}

The misaligned NS also leads to non-trivial polarization structure. This is because the electric field of the radio emission is aligned with the tangential component $B_t$ of the NS's magnetic field, but the direction of $B_t$ changes as a function of $\Omega t$. A more careful analysis is needed to verify if the polarization structure survives Faraday rotation induced from the magnetosphere, but we neglect this effect for now and study the polarization of the photons as they are emitted.  We define the basis vectors ${\bf \hat \epsilon_1} = {\bf \hat \theta}$ and ${\bf \hat \epsilon_2} = {\bf \hat \phi}$ for the polarization, where ${\bf \hat \theta}$ and ${\bf \hat \phi}$ are the polar and azimuthal unit vectors in the NS's frame.  Without loss of generality, we may consider the Earth to be at $\phi = 0$ in the frame of the NS, such that ${\bf \hat m} \cdot {\bf \hat r}$ is as given in~\eqref{mdr}.
The time-dependent polarization vector ${\bf \hat n}(t)$ of the radio emission is given by 
\es{}{
{\bf \hat n}(t) = {\big( \cos \theta_m \sin \theta -  \sin \theta_m \cos \theta \cos(\Omega t) \big) {\bf \hat \epsilon_1} - \sin \theta_m \sin(\Omega t) {\bf \hat \epsilon_2} \over \sqrt{ \big( \cos \theta_m \sin \theta - \sin \theta_m \cos \theta \cos(\Omega t)  \big)^2 + \sin^2 \theta_m \sin^2(\Omega t) }} \,.
}
In Fig.~\ref{fig: polarization}, we illustrate the linear polarization profiles, neglecting the possible effect of Faraday rotation, in the ${\bf \hat \epsilon_1}$ and ${\bf \hat \epsilon_2}$ directions over a pulsar period. Note that for $\theta = 90^\circ$ and a small misalignment angle $\theta_m = 15^\circ$, the outgoing wave is almost completely polarized over the whole rotation period.

\section{Radio sensitivity estimates}
In this section we give additional details for how we make the radio sensitivity estimates in the main Letter.

\subsection{Bandwidth optimization}

Considering that the asymptotic DM velocities, which determine the spread of photon frequencies, have a dispersion $\sim$$v_0$, it follows that we expect the bandwidth $B$ of the radio signal to be $B \sim m_a v_0^2 / (2 \pi)$, where the $2 \pi$ simply arises from the translation from angular frequency.  However, it is worth calculating the optimal bandwidth for an analysis of the radio-telescope data given that we are trying to optimize $\text{SNR} \sim \sqrt{B} S  \sim  F / \sqrt{B}$.  Having a smaller bandwidth increase our signal-to-noise ratio because of the $\sqrt{B}$ in the denominator, but on the other hand this also means that we have less flux $F$, which acts to decrease the signal. In this section, we compute the optimal bandwidth by maximizing $F /\sqrt{B}$.

Let us define 
\es{I_deff}{
I(\omega_1, \omega_2) \equiv {\int_{\omega_1}^{\omega_2} d \omega f_\omega(\omega) \over ([\omega_2 - \omega_1] / 2 \pi)^{1/2}} \,,
}
where $I(\omega_1, \omega_2)$ is the function to maximize and where $f_\omega(\omega)$ is the distribution of radio frequencies in the lab-frame for an axion signal with mass $m_a$.  In the DM frame, the DM velocity distribution is given by~\eqref{f_infty}.  Let us assume that the NS is boosted with a velocity ${\bf v_\text{b}}$ with respect to the DM frame.  Then, the distribution of frequencies in the NS frame, but asymptotically far away from the NS, is given by 
\es{f_omega}{
f_\omega(\omega) = {2 \over \sqrt{\pi} m_a v_b v_0} e^{ {2 - v_b^2 - 2 \tilde \omega \over v_0^2}} \sinh\left( {2 v_b \sqrt{2 \tilde \omega - 2} \over v_0^2} \right) \,,
} 
where we have defined $\tilde \omega \equiv \omega / m_a$.   This implies that 
\es{I_2}{
I(\omega_1, \omega_2) &= \sqrt{2 \pi \over m_a v_0^2} {1 \over \tilde B^{1/2} } {1 \over 2}\left[-2 + \text{erf}\big(x - \sqrt{2} \sqrt{\delta \tilde \omega_1} \big) + \text{erf}\big(x + \sqrt{2} \sqrt{\delta \tilde \omega_2} \big) + \text{erfc}\big(x + \sqrt{2} \sqrt{\delta \tilde \omega_1} \big) \right.\\
 &\left.+ \text{erfc}\big(x - \sqrt{2} \sqrt{\delta \tilde \omega_2} \big)   + 2 e^{ - x^2} {e^{-2 \delta \tilde \omega_1}\sinh(2 \sqrt{2 \delta \tilde \omega_1} x) - e^{-2 \delta \tilde \omega_2}\sinh(2 \sqrt{2 \delta \tilde \omega_2} x) \over \sqrt{\pi} x } \right] \,,
} 
with $x \equiv v_b /v_0$ and $\omega_{1,2} = m_a + m_a \delta \tilde \omega_{1,2} v_0^2$.  We have also defined $B \equiv (\omega_2 - \omega_1) / 2 \pi$ and $\tilde B \equiv 2 \pi B / (m_a v_0^2)$.  For $x=0$, we find the maximum at $\tilde \delta \omega_1 = 0.013$ and $\tilde B^* \approx 1.12$, with value $I(w_1, w_2) \sqrt{m_a v_0^2 / (2 \pi)} \approx 0.74$.  For $x = 1$, we find the maximum at $\tilde \delta \omega_1 = 0.032$ and $\tilde B^* \approx 1.97$, with value $I(w_1, w_2) \sqrt{m_a v_0^2 / (2 \pi)} \approx 0.58$.

Some explanation is warranted here.  First, we have not accounted for the gravitational potential of the NS for the reason mention in the main Letter: a DM wave with energy $\omega$ asymptotically far away from the NS will produce an electromagnetic wave of frequency $\omega$ when measured asymptotically far from the NS, since both the axion-photon mixing equations and the NS's gravitational potential conserve energy.  Second, we have not accounted for the boost of the lab frame with respect to the NS because this boost only affects the radio emission; such small boosts are not important for relativistic particles. On the other hand, the NS boost with respect to the DM distribution is important because DM is non-relativistic.

\subsection{Time-dependent light curves}  

Any analysis of real radio data for evidence of a DM signal will likely proceed through the use of a non-trivial likelihood function that properly accounts for the expected statistics of the observable, which may be the power within a frequency bin, under the null and signal hypotheses.  However, if we assume that the power measurements are normally distributed, we may write down a simple chi-square statistic that allows us to quickly estimate the sensitivity to axion signals with non-trivial light curves.  In particular, we may write 
\es{chi_square}{
\chi^2 = \sum_{i = 1}^{N_f} \sum_{j=1}^{N_t} { (S_{i,j} - S_b)^2 \over \sigma_{S_b}^2 } \,,
}
where $i$ labels the different independent frequency bins and $j$ labels the independent time bins.  Here, $S_{i,j}$ is the measured flux density within bin $(i,j)$, and $S_b$ is the predicted mean background flux density under the null hypothesis.  Note that we assume that $S_b$ is independent of $(i,j)$, though including dependence on the frequency and on time does not modify the main conclusion of this section.  Similarly, $\sigma_{S_b}^2$ is the expected variance of the flux density in a single bin under the null hypothesis.  Assuming the null hypothesis arises from thermal noise in the telescope, we may write 
\es{S_b}{
\sigma_{S_b} = { \text{SEFD} \over \sqrt{n_\text{pol} df dt} } \,,
}
where $df$ ($dt$) is the frequency (time) spacing between frequency (time) bins.

To estimate the mean expected significance of an axion signal, we follow the Asimov framework~\cite{Cowan:2010js} and take the data to be equal to the mean dataset under the signal hypothesis: $S_{i,j} = S_b + S^\text{axion}_{i,j}$, where $S^\text{axion}_{i,j}$ is the mean contribution from the axion in bin $(i,j)$.  Then, we find 
\es{}{
\chi^2_\text{Asimov} ={ n_\text{pol}  \over  \text{SEFD}^2}  \sum_{i = 1}^{N_f} df \sum_{j=1}^{N_t} dt  \,({{S^\text{axion}_{i,j}})^2} \,.
}
The signal will, roughly, have support over a range of frequencies of width equal to the bandwidth $B$, discussed in the previous subsection, so that we may write 
\es{}{
\chi^2_\text{Asimov} ={ n_\text{pol} B  \over  \text{SEFD}^2} \sum_{j=1}^{N_t} dt  \, ({{S^\text{axion}_{j}})^2  } \,.
}
A more careful likelihood function that includes the expected line-shape of the signal may be slightly more sensitive to a putative axion signal, but the development of such a likelihood is beyond the scope of the current work and not the focus of the current subsection.  We may write 
\es{}{
S^\text{axion}_{j} = S_0 \times f(j \times dt) \,,
}
where
\es{}{
S_0 = 6.7 \times 10^{-5} \, \, \text{Jy} \, \left({100 \, \, \text{pc} \over d} \right)^2  \left( {1 \, \, \text{GHz} \over m_a} \right)  \left({ 200 \, \, \text{km}/\text{s} \over v_0} \right)^2\left[\frac{d\mathcal{P} (\theta ={ \pi \over 2}, \theta_m = 0)/d\Omega}{4.5 \times 10^{8} \ {\rm W}}\right] \,,
}
and 
\es{}{
f(t) = \frac{3 \, ({\bf \hat m} \cdot {\bf \hat r})^2 + 1}{\big|3 \cos \theta \, {\bf \hat m} \cdot {\bf \hat r} - \cos \theta_m \big|^{4/3}} \Theta(r_c(t) - r_0)
}
with $\mathbf{\hat{m} \cdot \hat{r}}$ and $r_c(t)$ functions of time.  Then, we see that 
\es{Asimov}{
\chi^2_\text{Asimov} &={ n_\text{pol} B  \over  \text{SEFD}^2} S_0^2 \int_0^{\Delta t_\text{obs}} dt \, [f(t)]^2 \\
&={ n_\text{pol} B \Delta t_\text{obs}  \over  \text{SEFD}^2} {S_0^2} {\int_0^{2 \pi} d\phi \, [f(\phi)]^2 \over 2 \pi} \,, 
}
where $\Delta t_\text{obs}$ is the total time the radio telescope observed, and $\phi$ is the azimuthal coordinate on the NS.  That is,
\es{}{
{\bf \hat m} \cdot {\bf \hat r} & = \cos \theta_m \cos \theta + \sin \theta_m \sin \theta \cos(\phi) \,, \\
r_c(\phi) & = 224 \, \, \text{km} \times \left( {r_0 \over 10 \, \, \text{km}} \right) \left[ {B_0 \over 10^{14} \, \, {\rm G} } \ { 1\, \, \text{sec} \over P} \left( {1 \, \, \text{GHz} \over m_a} \right)^2 \right]^{1/3}  | 3 \cos \theta \, {\bf \hat m} \cdot {\bf \hat r} - \cos \theta_m |^{1/3} \,,
}
where we can also identify $\phi = \Omega t$.  Evaluating~\eqref{Asimov} and solving for $g_{a \gamma \gamma}$ such that $\chi^2_\text{Asimov} = 1$ leads to the projected sensitivities in Fig.~\ref{fig: limits}.

\subsection{Angular dependence of sensitivity}

\begin{figure}[t!]
\includegraphics[width = 0.5\textwidth]{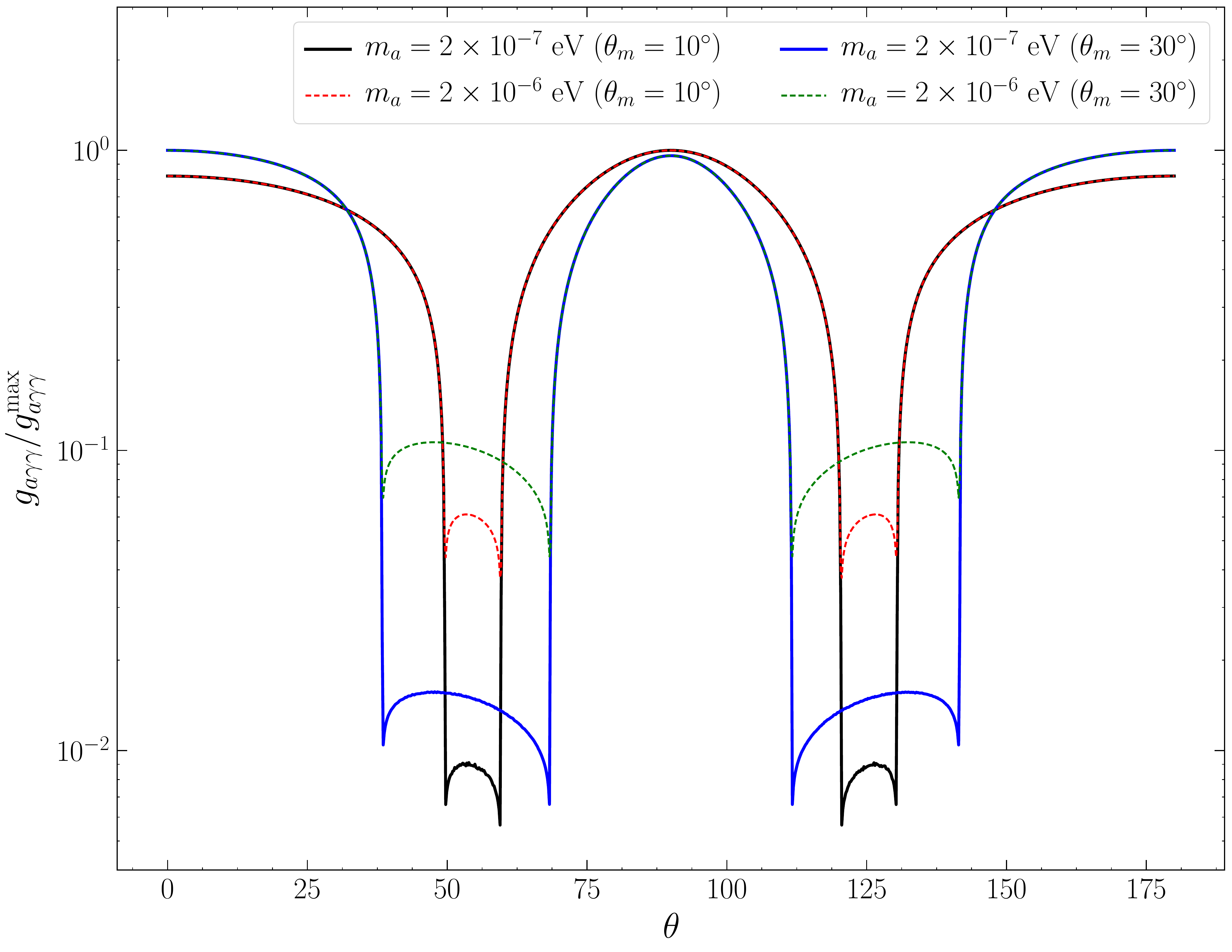}
\caption{ Sensitivity to $\g$ for two different axion masses $m_a$ and two different misalignment angles $\theta_m$, as a function of $\theta$, the polar angle of the Earth in the frame of the NS. The remaining parameters, chosen for illustration, are $B_0 = 2.5 \times 10^{13} \ {\rm G}$, $P = 1~{\rm s}$, and $r_0 = 10~{\rm km}$. Smaller values of $\g/\g^{\rm max}$ indicate an enhancement.}
\label{fig: comparison}
\end{figure}

In Fig.~\ref{fig: limits} we showed the projected sensitivity to $g_{a \gamma \gamma}$ as a function of mass for two different angles $\theta$, with $\theta_m = 10^\circ$.  However, it is important to understand how the sensitivity changes a function of $\theta$, given that in practice this will be a random angle that depends on the relative orientation of the Earth and the NS being observed.  In Fig.~\ref{fig: comparison} we show how the sensitivity changes as a function of $\theta$ for the two cases $\theta_m = 10^\circ,~30^\circ$. We illustrate two different masses, $m_a = 2 \times 10^{-7}$ eV and $m_a = 2 \times 10^{-6}$ eV, and normalize the sensitivity to the maximum (worst) value, $g_{a \gamma \gamma}^\text{max}$, found over all $\theta$.  Note that the large enhancement due to the strong beaming is found over a relatively wide range of $\theta$, so we may expect such an enhancement for generic NS targets.

\end{document}